\documentclass{jltp}
\usepackage{graphicx}
\usepackage{psfrag}
\usepackage{overcite}

\begin{document}

\title{Dispersive Charge and Flux
Qubit Readout as a Quantum Measurement Process }

\author{Lars Tornberg and G\"oran Johansson}

\address{Applied Quantum Physics, MC2, Chalmers, S-412 96
G\"oteborg, Sweden}

\runninghead{L. Tornberg and G. Johansson}{Quantum efficiency of
charge and flux qubit readout}

\maketitle

\begin{abstract}
We analyze the dispersive readout of superconducting charge and
flux qubits as a quantum measurement process. The measurement
oscillator frequency is considered much lower than the qubit
frequency. This regime is interesting because large detuning
allows for strong coupling between the measurement oscillator and
the signal transmission line, thus allowing for fast readout. Due
to the large detuning we may not use the rotating wave
approximation in the oscillator-qubit coupling. Instead we start
from an approximation where the qubit follows the oscillator
adiabatically, and show that non-adiabatic corrections are small.
We find analytic expressions for the measurement time, as well as
for the back-action, both while measuring and in the off-state.
The quantum efficiency is found to be unity within our
approximation, both for charge and flux qubit readout.
\end{abstract}


\section{INTRODUCTION}
Superconducting qubits are considered one of the technologies
capable of bringing us quantum information processing (QIP) devices
in the future\cite{WendinShumeiko2005}. One cornerstone of QIP is
the possibility to convert the quantum state into classical
information, i.e. read-out. A large effort has gone into optimizing
different read-out techniques, see e.g. Ref.
\onlinecite{JohanssonJPCM2006} and references therein. Of course
reading out a qubit only makes sense if it can stay coherent long
enough to contain true quantum information. Being solid-state
qubits, superconducting qubits interact rather strongly with their
environment, making dephasing the worst enemy of superconducting
QIP.

Quite naturally, which type of noise is most dangerous depends on
the qubit design. For example charge qubits\cite{nakamura} are
most sensitive to charge fluctuations\cite{AstafievPRL2004}, while
flux qubits are most sensitive to flux
noise\cite{ChiorescuScience2003} and charge-phase
qubits\cite{VionScience2002} are sensitive to
both\cite{IthierPRB2005}. At the degeneracy point the two
charge/flux qubit states have equal average charge/flux. Thus, at
this point they are protected against environmental charge/flux
noise to first order in the coupling. To minimize the effect of
the typical noise, qubits are now operated so that they remain at
this degeneracy point as much as possible.

The natural readout for a charge/flux qubit is to measure the
average charge/flux of the qubit. Since the qubit states have
equal average charge/flux at the degeneracy point, the natural
type of readout is no longer possible. This initiated an interest
in dispersive readout techniques, where the qubit state influences
the properties of a harmonic oscillator, which is then
probed\cite{GrajcarPRB04,LupascuPRL04,WallraffPRL05}.

In the experiment of Wallraff {\it et. al.} at
Yale\cite{WallraffPRL05}, a charge qubit was coupled to an
oscillator, realized by a coplanar waveguide. Using this setup
they managed to determine the state of the qubit, while
maintaining long coherence times. In this case, the oscillator and
qubit where nearly degenerate in frequency and an oscillator with
a high quality factor $Q$ was used to shield the qubit from
relaxation.

In this paper, we concentrate on the situation where the qubit and
oscillator frequencies differ by orders of magnitude. This will
allow us to use a low $Q$ oscillator, thereby increasing the
read-out speed of the circuit. Here it is the detuning that
shields the qubit from measurement induced mixing, and also allows
relaxation to be minimized through effective filtering of the
transmission line at the qubit frequency. In this setup the state
of a charge/phase qubit can be read out through its effective
capacitance/inductance, which shifts the resonant frequency. At
the degeneracy point, the state dependent quantum
capacitance/inductance for the two states have the same magnitude
but differ in sign. For the charge qubit the quantum capacitance
was recently measured by two groups \cite{DutyPRL,Sillanpaa}.

The theory of quantum measurement states the impossibility to
discern the state of a quantum system without destroying the phase
coherence in the system. This promotes the idea of defining the
quantum efficiency $\eta$ of a measurement as the ratio between
the dephasing time $t_\varphi$ and measurement time $t_{ms}$,
bounded above by $\eta = t_\varphi/2t_{ms} \leq
1$\cite{Braginsky}. (To be consistent with the results of the Yale
group \cite{schoel}, we have decided to use their definition of
measurement time. This, together with the desire to keep $\eta
\leq 1$ is the reason for the appearance of the one half in the
definition). The quantum efficiency gives a measure on the
back-action of the read-out and deriving $\eta$ in terms of
circuit parameters is thus a good starting point for optimizing
the read-out.

In Ref.~\onlinecite{JohanssonTornbergWilson}, together with C. M.
Wilson, we investigated the quantum efficiency of the read-out
proposal mentioned above, as well as discussed the performance of
a realistic circuit design. Here we give the more detailed
theoretical background for those results. The large detuning
between the qubit and oscillator makes standard methods of quantum
optics \cite{zoller,schoel} inappropriate. Instead, we start from
an approximation where the qubit follows the oscillator
adiabatically, and show that non-adiabatic corrections are small.
Our results are based on the quantum network theory, introduced by
Yurke and Denker\cite{YurkeDenkerPRA}. As already discussed in
Ref.~\onlinecite{JohanssonTornbergWilson}, the efficiency for both
charge and phase qubit readout is ideal within our approximation.
This result is independent of the oscillator $Q$-value.
Furthermore, we find that a low $Q$ oscillator shields the qubit
from thermal dephasing. Finally we treat explicitly the limiting
case of a transmission line capacitively coupled directly to a
charge qubit. Being of more academic interest, we find an ideal
quantum efficiency also in this case.

\section{CHARGE QUBIT}\label{sec:lagrange}
The goal of this section is to derive the Hamiltonian for the
superconducting charge qubit complete with read-out device. Having
the full quantum description of the circuit we then make suitable
approximations to derive the quantum capacitance of the
Cooper-pair box. This gives us an adiabatic Hamiltonian which we
can solve to obtain the dynamics of the system exactly. In section
\ref{sec:LZ}, we give bounds for the non-adiabatic contributions,
and justify our approximations in detail. In the rest of the paper
we use the adiabatic Hamiltonian to investigate the efficiency of
our
measurement scheme. \\
\begin{figure}[!ht]
\psfrag{Vg}[][][0.9]{$V_g$}
\psfrag{phi_g}[][][0.9]{$$}
\psfrag{C_g}[][][0.9]{$C_g$}

\psfrag{phi_m}[][][0.9]{$$}
\psfrag{C_m}[][][0.9]{$C_m$}

\psfrag{phi_J}[][][0.9]{$$}
\psfrag{C_J}[][][0.9]{$C_J$}
\psfrag{E_J}[][][0.9]{$E_J$}

\psfrag{phi_l}[][][0.9]{$$}
\psfrag{L}[][][0.9]{$L$}

\psfrag{phi_C}[][][0.9]{$$}
\psfrag{C}[][][0.9]{$C$}

\psfrag{phi_in}[][][0.9]{$$}
\psfrag{C_c}[][][0.9]{$C_c$}

\psfrag{phi1}[][][0.9]{$$}
\psfrag{phi2}[][][0.9]{$$}
\psfrag{C_c}[][][0.9]{$C_c$}
\psfrag{C_T}[][][0.9]{$C_T\Delta x$}
\psfrag{L_T}[][][0.9]{$L_T\Delta x$}

\psfrag{deltaphi}[][][0.9]{$$}

\psfrag{ChargeQubit}[][][0.9]{Charge Qubit}
\psfrag{Oscillator}[][][0.9]{Oscillator}
\psfrag{TransmissionLine}[][][0.9]{Transmisson Line}

\includegraphics[width=12cm]{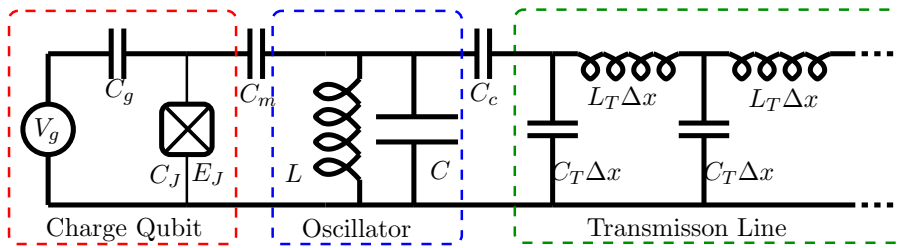}
\caption{Cooper-pair box operated as a charge-qubit complete with
  read-out circuitry, see text.}
\label{fig:circuitCharge}
\end{figure}
The single Cooper-pair box operated as a charge
qubit\cite{ShnirmanSCQPRL} can be viewed in Fig.
\ref{fig:circuitCharge}. A small superonducting island is
connected to a superconducting lead through a Josephson junction
with Josephson energy $E_J$ and capacitance $C_J$. The island is
also coupled to a gate voltage $V_g$ via the gate capacitance
$C_g$. By tuning $V_g$, the desired working point of the qubit can
be reached. To enable readout an LC-oscillator is coupled to the
box via a measuring capacitance $C_m$. The oscillator is in turn
coupled to a transmission line, here modelled as a semi-infinite
line of LC-oscillators in parallel. The line is characterized by
its capacitance $C_T$ and inductance $L_T$ per unit length.
Through this line, all measurements of the qubit will be
performed.

We choose, as our starting point, to write down the Lagrangian for
the system \cite{YurkeDenkerPRA,devoret}. This method gives
automatically the conjugated variables of the system and thus
provides us with the dynamics. Since the circuit contains a
Josephson element the natural choice of coordinates is the phase
across each circuit element, which is related to the voltage drop
by $\phi_i = \int V_i dt$. Following standard
procedure\cite{devoret} the capacitive and inductive energy of the
circuit are the kinetic and potential energy of the Lagrangian
\begin{eqnarray}\label{eq:T}
T &=& \frac{C_g \dot{\Phi}^2_g}{2} +
    \frac{C_J \dot{\Phi}^2_J}{2} +
    \frac{C_m \dot{\Phi}^2_m}{2} +
    \frac{C \dot{\Phi}_C^2}{2} +
    \frac{C_c \dot{\Phi}^2_{in}}{2}
+  \sum_{i=1}^\infty\Delta x \frac{C_T( \dot{\Phi}^p_{i}
    )^2}{2}\nonumber \\
V &=& \frac{\Phi_L^2}{2 L} - E_J
\cos\left(\frac{\Phi_J}{\Phi_0}\right) + \sum_i\Delta x
\frac{(\Phi^p_{i+1} - \Phi^p_{i} )^2}{2L_T(\Delta x)^2},
\end{eqnarray}
where $\Phi_0 = \hbar/2e$ is the flux quantum.
Applying  Kirchoff's voltage law gives the constraints on the circuit
coordinates
which gives the Lagrangian, $L = T - V$, for the system
\begin{eqnarray}\label{eq:lagrangian2}
L &=& \frac{C_{qb} \dot{\Phi}^2_J}{2} +
\frac{\left(C_{osc}+C_c\right) \dot{\Phi}_C^2}{2} + \frac{C_c(
\dot{\Phi}^p_1 )^2}{2} - \frac{\Phi^2_C}{2 L} + E_J
\cos\left(\frac{\Phi_J}{\Phi_0}\right) - \nonumber \\
&-& C_m\dot{\Phi}_C\dot{\Phi}_J - C_c \dot{\Phi}_C\dot{\Phi}_1^p -
C_gV_g\dot{\Phi}_J + \frac{C_g V_g^2}{2} +\nonumber\\
 &+& \sum_{i=1}^\infty \Delta x \left( \frac{C_T( \dot{\Phi}^{p}_i )^2}{2} -
      \frac{(\Phi^{p}_{i+1} - \Phi^{p}_{i} )^2}{2L_T(\Delta x)^2}\right),
\end{eqnarray}
where we have introduced the capacitances  $C_{qb} = C_J + C_g +
C_m$, and $C_{osc} = C + C_m$ for brevity. To simplify the
derivation of the Hamiltonian we introduce the vector notation for
our coordinates
\begin{equation}
\vec{\Phi} = \left[
\begin{array}{ccccc}
\Phi_J &
\Phi_C &
\Phi_1^p &
\Phi_2^p &
\ldots
\end{array}
\right].
\end{equation}
The Lagrangian can now be written more compactly
\begin{equation}\label{eq:Lagrangian_vec}
L = \frac{1}{2}\dot{\vec{\Phi}}^T C \dot{\vec{\Phi}} -
C_gV_g\dot{\Phi}_J  -  \frac{\Phi_C^2}{2 L} +
E_J \cos\left(\frac{\Phi_J}{\Phi_0}\right) - \sum_i\Delta x  \frac{(\Phi^p_{i+1} - \Phi^p_{i}
)^2}{2L_T(\Delta x)^2},
\end{equation}
 where $C$ is the mass matrix of the system. This can be divided into
 $C_{sys}$, describing the capacitative elements in the
 qubit-oscillator circuit and a pure diagonal matrix $C_{TL}$ which
 only contains information about the capacitance per unit length in
 the transmission line
\begin{eqnarray}
C & = & \left[
\begin{array}{c|c}
C_{sys} & 0 \\
\hline 0  & C_{TL}
\end{array}
\right] , \qquad
C_{sys} = \left[
\begin{array}{ccccc}
C_{qb} & -C_m & 0   \\
-C_m & C_{osc}+C_c& -C_c \\
0 & -C_c & C_c+C_T \Delta x   \\
\end{array}
\right] , \nonumber \\
C_{TL} &=& \left[
\begin{array}{ccc}
C_T & 0 & \ldots \\
0 & C_T & \ddots \\
\vdots & \ddots & \ddots
\end{array}
\right].
\end{eqnarray}
The Lagrangian contains all information needed for deriving the
system dynamics. However, when doing quantum mechanical
calculations one often prefers to work with the Hamiltonian. We
thus Legendre transform\cite{goldstein} the Lagrangian following
 \ref{legendre} and arrive at
\begin{equation}\label{eq:H_tot}
H = H_{qb} + H_{osc} + H_{int} + H_{TL},
\end{equation}
where $H_{qb}$ describes the Josephson junction alone and also the
interaction between JJ and oscillator
\begin{equation}\label{eq:Hq}
H_{qb} =  \frac{D_{11}}{2}(p_J + C_gV_g)^2 + \left( D_{12}p_C +
D_{13}p_p \right)(p_J + C_gV_g) -
E_J\cos\left(\frac{\Phi_J}{\Phi_0}\right),
\end{equation}
and where $D_{ij}$ are the elements of the matrix
\begin{equation}
D=C_{sys}^{-1}=\frac{1}{C_{qb}C_{osc}-C_m^2} \left[
\begin{array}{ccccc}
C_{osc} & C_m & C_m   \\
C_m & C_{qb}& C_{qb} \\
C_m & C_{qb} & \left(C_{qb}C_{osc}-C_m^2\right)/C_c+C_{qb} \\
\end{array}
\right].
\end{equation}
The last three terms in the Hamiltonian, Eq. (\ref{eq:H_tot}),
describe the oscillator, the transmission line and the coupling
between them;
\begin{eqnarray}\label{eq:Hrest}
H_{osc} &=&  \frac{D_{22}}{2}p_C^2 + \frac{\Phi^2_C}{2L}, \qquad
\nonumber H_{int} = D_{23}p_C p_p , \nonumber \\
H_{TL} &=&  \frac{D_{33}}{2}p_p^2 + \frac{1}{\Delta x}
\sum_{i=1}^\infty \left( \frac{ (p_{(i+1)}^p)^2}{2C_T} +
\frac{(\Phi^p_{i+1} - \Phi^p_{i} )^2}{2L_T}\right),
\end{eqnarray}
where the operators $p_C, p_J, p_p$ and $p_{i}^p$ are the
conjugate momenta to $\Phi_C, \Phi_{J}, \Phi^p_{1}$ and
$\Phi^p_{i}$ respectively, and have dimension of charge. By
definition, the phases and charges $\phi_i$ and $p_j$ obey the
canonical commutations relation $[\phi_i,p_j] =
i\hbar\delta_{ij}$. The kinetic energy of the Hamiltonian is thus
represented by the the charging energies of the circuit. The
$D$-matrix contains information about the effective capacitances
associated with the respective charges.

The JJ island is caracterized by a charging energy, $E_C =
D_{11}e^2/2 \approx e^2/2C_{qb}$ and the Josephson energy $E_J$. To
realize a two level system, the charging energy of the island needs
to be much larger than the Josephson energy. For this choice of
parameters we can limit the number of excess electrons on the island
to the values $\{0,2\}$. Projecting the qubit Hamiltonian on this
subspace of the whole Hilbert space in charge basis, we get the
usual qubit Hamiltonian in the language of Pauli spin matrices
\begin{equation}\label{eq:Hq3}
H_{qb} =  -\frac{E_{el}}{2}\sigma_z - \frac{E_J}{2}\sigma_x  +
2E_c\kappa(n_C + n_p)(1-n_0) + E_C (2-n_0^2),
\end{equation}
where $\kappa = D_{12}/D_{11} = C_m/C_{osc}$ is a dimensionless
coupling constant and $E_{el} = 4E_c(1-\tilde{n})$ is the
electrostatic energy of the island depending on both gate charge
and induced charge from the oscillator and transmission line
$\tilde{n} = n_0 + \kappa( n_C + n_p)$. The number operators $n$
are related to the charges by $n_C = p_C/e$, $n_p = p_p/e$  and $
n_0 = C_gV_g/e$ respectively. The last term in Eq. (\ref{eq:Hq3})
is just an offset in the energy of the system and thus discarded.

\subsection{QUANTUM CAPACITANCE}
During read-out the oscillator will be driven close to the bare
resonance frequency which we consider much smaller than the qubit
level-splitting $E_J/\hbar$. This difference in frequency allows
us to make an adiabatic approximation, separating the qubit and
oscillator dynamics. Thus the qubit eigenvalues and eigenstates
will depend on the oscillator and transmission line coordinates.
However the levels will never cross and the qubit is frozen in its
state. In this regime, the qubit can be described by an effective
adiabatic Hamiltonian
\begin{equation}\label{eq:Hq4}
H_{\rm A} = \frac{\sqrt{E_{el}(\tilde{n})^2 + E_J^2}}{2}\sigma_z +
2E_c\kappa (n_C+n_p)(1-n_0),
\end{equation}
which is obtained by a unitary transformation of the the qubit
Hamiltonian in Eq. (\ref{eq:Hq3}). For more details on this, see
Sec. \ref{sec:LZ} The expression in Eq. (\ref{eq:Hq4}) is similar
to the usual expression for the Hamiltonian of a charge qubit in
computational basis. The difference in this case is that the
working point of the qubit is determined not only by the $V_g$
dc-bias, but also on the induced charge on the oscillator. The
second term is usually proportional to the identity operator and
ignored. Here it contains both system and transmission line
operators, and can not be thrown away. However, it only represents
a small shift in $p_C$ and $p_1$ and can thus be included into a
new set of shifted operators. Considering a small coupling between
qubit and oscillator we can expand the qubit energy to second
order in $\delta n = \kappa (n_C + n_p)$. The effect of the qubit
can thus be summarized in three terms; one state dependent off-set
which does not affect the circuit dynamics, a linear term in
$\kappa (n_C + n_p)$ which is zero at the degeneracy point, and
finally a term quadratic in oscillator and transmission line
charge. Discarding the two first we arrive at the Hamiltonian
determining the low frequency dynamics of the oscillator and
transmission line, at the charge degeneracy point
\begin{eqnarray}\label{eq:H_charge}
H &=& \frac{D_{22}}{2} p_C^2  + \frac{\Phi_C^2}{2L} +
\frac{g_C}{2}\sigma_z (p_C+p_p)^2 \nonumber\\
 & + & \frac{D_{33}}{2}p_p^2 + \frac{1}{\Delta x} \sum_i
\left( \frac{ (p_{(i+1)}^p)^2}{2C_T} + \frac{(\Phi^p_{i+1} -
\Phi^p_{i} )^2}{2L_T}\right) + D_{23} p_C p_p.
\end{eqnarray}
Here we have defined the coupling between the charge qubit and
oscillator $g_C = C_Q/C_{osc}^2 = \frac{8E_c^2\kappa^2}{e^2E_J} $,
and the \emph{quantum capacitance} as
\begin{equation}
C_Q = -\frac{2e^2C_m^2}{E_J C_{qb}^2} \sigma_z.
\end{equation}
The motivation behind this definition comes from the fact that the
qubit shifts the electrostatic energy of the oscillator, which can
be interpreted as a shift of the oscillator capacitance, $C_{osc}
\rightarrow C_{osc} + C_Q$. This interpretation is possible as long
as $C_Q \ll C_{osc}$. The corresponding qubit induced shift of the
oscillators resonance frequency is simply given by $\omega_{osc} =
1/\sqrt{L(C_{osc}+C_Q)}$.

\section{FLUX QUBIT AND QUANTUM INDUCTANCE}
\begin{figure}[!ht]
\psfrag{phi_g}[][][0.9]{$$}
\psfrag{C_g}[][][0.9]{$C_g$}

\psfrag{phi_m}[][][0.9]{$$}
\psfrag{C_m}[][][0.9]{$C_m$}

\psfrag{phi_J}[][][0.9]{$$}
\psfrag{C_J}[][][0.9]{$C_J$}
\psfrag{E_J}[][][0.9]{$E_J$}

\psfrag{phi_l}[][][0.9]{$$}
\psfrag{L}[][][0.9]{$L$}

\psfrag{phi_C}[][][0.9]{$$}
\psfrag{C}[][][0.9]{$C$}

\psfrag{phi_in}[][][0.9]{$$}
\psfrag{C_c}[][][0.9]{$C_{in}$}

\psfrag{phi1}[][][0.9]{$$}
\psfrag{phi2}[][][0.9]{$$}
\psfrag{C_c}[][][0.9]{$C_{in}$}
\psfrag{C_T}[][][0.9]{$C_T\Delta x$}
\psfrag{L_T}[][][0.9]{$L_T\Delta x$}

\psfrag{deltaphi}[][][0.9]{$$}

\psfrag{PHI}[][][1.5]{$\Phi_x$} \psfrag{M}[][][0.9]{$M$}
\psfrag{L_J}[][][0.9]{$L_J$} \psfrag{PhaseQubit}[][][0.9]{Flux
Qubit} \psfrag{Oscillator}[][][0.9]{Oscillator}
\psfrag{TransmissionLine}[][][0.9]{Transmission Line}

\includegraphics[width=12cm]{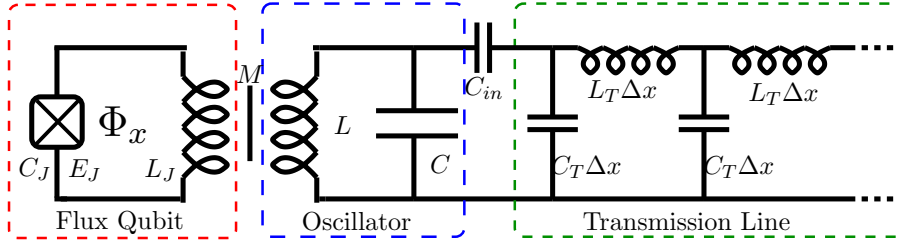}
\caption{Radio-frequency SQUID operated as a flux-qubit, complete
with
  read-out circuitry, see text.}
\label{fig:circuitFlux}
\end{figure}

The radio-frequency superconducting quantum interference device
(rf-SQUID) operated as a flux qubit\cite{FriedmanNature2000} can
be seen in Fig. \ref{fig:circuitFlux}. The readout part of our
analysis also applies to the 3-junction persistent current
qubit\cite{ChiorescuScience2003}, but the rf-SQUID is chosen for
simplicity. In the previous section the charge states of the
single Cooper-pair box were used as the computational basis of the
qubit. Here the qubit loop is biased by an external flux $\Phi_x$,
and the $|0\rangle$ and $|1\rangle$ states are realized by the
clock and anticlockwise circulating supercurrents in the loop. The
read-out circuit is the same as for the charge qubit. It is
however inductively coupled to the qubit via the mutual inductance
$M$, as opposed to the capacitative coupling in the charge case.
The derivation of the Hamiltonian for this circuit is almost
analogous to the analysis for the charge qubit. Making the adiabatic
approximation and rotating to the qubit eigenbasis yields the
following adiabatic Hamiltonian for the oscillator and transmission
line, considering the qubit biased at its degeneracy point
\begin{eqnarray}\label{eq:H_flux}
H &=& \frac{p^2}{2C} + \left( \frac{1}{2L} -
\frac{M^2\langle\phi\rangle^2}{\Delta(LL_J)^2}
\sigma_z\right)\Phi^2 + \frac{(p_p)^2}{2C_c} + \frac{p_pp}{C}
+ \nonumber\\
&+& \frac{1}{\Delta x} \sum_i \left( \frac{ (p_{(i+1)}^p)^2}{2C_T}
+ \frac{(\Phi^p_{i+1} - \Phi^p_{i} )^2}{2L_T}\right),
\end{eqnarray}
where $C_c = (C + C_{in})/(CC_{in})$, $\langle\phi\rangle$ is the
magnitude of the flux generated by the qubit persistent current in a
current eigenstate, and $\Delta$ is the qubit energy splitting. Just
as in the charge case, the slow transverse coupling to the
oscillator is seen as a longitudinal coupling in the eigenbasis of
the qubit. Due to the phase-coupling between oscillator and qubit we
get the coupling constant $g_L = L_Q/L^2$ where the quantum
inductance is defined as
\begin{equation}
L_Q = \frac{M^2\langle\phi\rangle^2}{\Delta L_J^2},
\end{equation}
so that the coupling can be incorporated into an effective
inductance $L^{g/e} \equiv L \pm L_Q$.

\section{NON-ADIABATIC CORRECTIONS AND QUBIT RELAXATION}\label{sec:LZ}
In the following we will calculate the dynamics of the oscillator
and transmission line, considering the qubit state to be fixed in
its adiabatic eigenbasis. This is of course an approximation, and
there are two different effects induced by the measurement circuit
that will act to change the qubit state. First, too strong driving
will induce transitions, which in the large amplitude limit will
be of Landau-Zener type. Secondly, seen from the qubit
perspective, the measurement circuit consists of a dissipative
transmission line filtered through the oscillator. This will add
to the relaxation rate of the qubit, also when not measuring.

\subsection{Non-Adiabatic Corrections}
Equation (\ref{eq:Hq4}) was derived by diagonalizing the qubit
Hamiltonian, a transformation which depends on the state of both
oscillator and transmission line. In the adiabatic regime however,
it is possible to find eigenvalues and eigenstates of the qubit
for each value of the charge on the oscillator and transmission
line. Because of the above, the transformation will however also
affect the oscillator part of the Hamiltonian, and we must
consider this effect and make sure that it is negligible. We begin
with the unrotated Hamiltonian for qubit and oscillator (see
Eq.~(\ref{eq:Hq3}))
\begin{equation}
H = -\frac{E_{el}(\tilde{n})}{2}\sigma_z - \frac{E_J}{2}\sigma_x +
\frac{p^2}{2C} + \frac{\Phi^2}{2L},
\end{equation}
where the coupling is in $E_{el}$, as assumed before. Making the
unitary transformation to qubit eigenbasis\cite{makhlin}
\begin{equation}
\tilde{H} = u^\dagger H u, \qquad u = \left(
\begin{array}{cc}
\cos\eta/2 & -\sin\eta/2 \\
\sin\eta/2 & \cos\eta/2
\end{array}\right), \qquad
\eta = \mathrm{arccot}\left(\frac{E_{el}(\tilde{n})}{E_J}\right)
\end{equation}
diagonalizes the qubit part of $H$, but also affects the
oscillator term that contains the phase. At the degeneracy point
$n_0 =1$ the transformation is given to first order in $\kappa$
\begin{equation}
u \simeq  \frac{1}{\sqrt{2}}\left[\left(
\begin{array}{cc}
1&1\\
-1&1
\end{array}\right) + \frac{x}{2}\left(
\begin{array}{cc}
1&-1\\
1&1
\end{array}\right)\right],
\end{equation}
where $x = 4E_c\kappa \tilde{n}/E_J$. The term that we have
disregarded when deriving Eq. (\ref{eq:Hq4}) is thus

\begin{eqnarray}\label{eq:throw}
u^\dagger \frac{\Phi^2}{2L}u &=&   \frac{1}{2}\left(
\begin{array}{cc}
0&1\\
-1&0
\end{array}\right)[x,\frac{\Phi^2}{2L}] = \frac{i\hbar}{2} \dot{x}\left(
\begin{array}{cc}
0&1\\
-1&0\end{array}\right),
\end{eqnarray}
which would induce transitions in the qubit. It must therefore be
compared with the natural transition energy in the qubit, given by
the level splitting $E_J$. Since we explicitly assume both small
amplitude $4\kappa \delta n E_c \ll E_J$ and low frequency $\hbar
\omega \ll E_J$ of the driving, the ratio is given by
\begin{equation}\label{eq:condition}
 \frac{4\kappa E_c \hbar\omega \delta n }{E_J}\frac{1}{E_J} \ll 1,
\end{equation}
which is indeed negligible within our approximation. We note the
similarity of the above condition and the one needed for
neglecting
 Landau-Zener(LZ) transitions\cite{LandauZener}. The LZ transition
 rate is given by
\begin{equation}
\Gamma_{LZ}\propto e^{-2\pi E_J^2/\hbar v}, \qquad v=\partial_t
E_{el},
\end{equation}
which thus also can be neglected if the condition in Eq.
(\ref{eq:condition}) is fulfilled. For the flux qubit the
corresponding condition for small amplitude driving is
\begin{equation}\label{eq:condition_flux}
\frac{M|\langle\phi\rangle| \Phi}{\Delta L_J L} \ll 1 .
\end{equation}
In Sec. \ref{sec:QuantEff} we will find that the quantum
efficiency is independent of the drive strength. But the speed of
the measurement is proportional to the power of the drive, as we
will see in Sec \ref{sec:readout}. Thus, in a real experiment one
would like to drive as strongly as possible, but one should then
make sure not to violate Eqs.~(\ref{eq:condition}) or
(\ref{eq:condition_flux}).

\subsection{Qubit Relaxation}
Although the {\em pure dephasing} induced by the transversely
coupled oscillator circuit vanishes to first order in the
oscillator-qubit coupling, the qubit {\em relaxation} rate
$\Gamma_1$ has a first order contribution. Thus we evaluate the
relaxation rate using standard weak coupling
expressions\cite{makhlin}, giving that it is proportional to the
real part of the impedance seen from the qubit, at the qubit
frequency $\omega_{qb}$. For the simple oscillator circuits shown
in Figs.~\ref{fig:circuitCharge} and ~\ref{fig:circuitFlux} with a
low $Q$, the relaxation induced by the transmission line can
easily dominate over other relaxation mechanisms. The solution to
this apparent problem is to add a non-dissipative low-pass filter
between the transmission line and the oscillator. Due to the large
frequency difference between the oscillator and qubit, this is
straightforward. The high-frequency electromagnetic environment
responsible for qubit relaxation can be designed independently of
the low frequency environment which should let the readout signal
through. In principle, a commercially available pi-filter can
improve the relaxation rate by a factor of 1000, while not
affecting the low frequency measurement properties. For realistic
designs\cite{JohanssonTornbergWilson} this is well beyond the
point where other sources of relaxation will dominate.

\section{SYSTEM DYNAMICS}
The questions that we address in this paper concern the efficiency
of read-out by probing the quantum capacitance (inductance) of a
charge (flux) qubit. The basic idea is that we send a pulse
through the transmission line, which will be reflected by the
combined oscillator-qubit circuit. In our approximation the pulse
will see two different harmonic oscillators, depending on the
qubit state. While the information about the state of the qubit
state thus is transferred to the reflected pulse, the unavoidable
photon number fluctuations in the pulse will dephase the qubit.
The statistics of these fluctuations is determined by the way the
pulse is generated, and the dynamics of the transmission line
coupled to a harmonic oscillator. We thus start out by calculating
the dynamics of the free transmission line. Given this, the
qubit-oscillator circuit will act as a boundary condition to this
solution.

\subsection{Equations of Motion: Transmission Line}\label{sec:eqnsofmot_TL}
The time-evolution for the transmission line degrees of freedom is
governed by the $H_{TL}$ in Eq. (\ref{eq:Hrest}). For $x > 0$ we
consider the transmission line phases and charges in the limit
$\Delta x\rightarrow 0$. In this limit the Heisenberg equations
yield the coupled equations, two for each $i>0$
\begin{eqnarray}
\dot{p}_i &=& \left[ p_i , \frac{\Delta x}{2L_T} \frac{(\Phi_{i+1}
-\Phi_{i})^2  - (\Phi_{i} -\Phi_{i-1})^2 }{(\Delta x)^2}  \right]
\rightarrow \frac{\mathrm{d}x}{L_T}\frac{\partial^2
\Phi_i}{\partial x^2}, \nonumber \\
p_i &=& C_T \mathrm{d} x \dot{\Phi}_i .
\end{eqnarray}
Combining these two equations to one we see that the phase in the
transmission lines obeys the massless scalar Klein-Gordon equation
\begin{equation}\label{eq:KleinGordon}
\frac{\partial^2\Phi}{\partial t^2} - \frac{1}{L_T
C_T}\frac{\partial^2 \Phi}{\partial x^2} = 0,
\end{equation}
which indicate the bosonic nature of the transmission line
excitations. Equation (\ref{eq:KleinGordon}) has a solution in the
form of right and leftgoing travelling waves which formally can be
written as
\begin{equation}\label{eq:KG_sol}
\Phi = \Phi^{in}\left( \frac{x}{v} + t \right) + \Phi^{out}\left(-
\frac{x}{v} + t \right),
\end{equation}
where $v=1/\sqrt{L_T C_T}$ is the velocity of the waves in the
transmission line. By differentiating Eq. (\ref{eq:KG_sol}) we can
get a relation between the partial
derivatives\cite{YurkeDenkerPRA}
\begin{equation}\label{eq:diff:kg_sol}
-\frac{1}{L_T}\frac{\partial \Phi}{\partial x} =
\sqrt{\frac{C_T}{L_T}}\left(  \frac{\partial \Phi}{\partial t} - 2
\frac{\partial \Phi^{in}}{\partial t}\right).
\end{equation}
The spatial derivative of the probing field at $x=0$ together with
Eq. (\ref{eq:diff:kg_sol}) will later provide us with the coupling
between circuit and transmission line. It is however natural to
first derive the statistics of the transmission line and then go
on to the circuit dynamics.

\subsection{Expressing the Transmission Line Using Creation and
  Annihilation Operators}\label{sec:createannihilate}

The Klein-Gordon equation of motion is easily solved by
introducing the Fourier transformed operators in the transmission
line
\begin{equation}
\Phi(x,t) =
\frac{1}{\sqrt{2\pi}}\int_{-\infty}^\infty\Phi(k,t)e^{ikx}dk,
\end{equation}
and we get an ordinary differential equation for each value of the
wave-number $k$. Solving Eq. (\ref{eq:KleinGordon}) is reduced to
the problem of solving
\begin{equation}
\ddot{\Phi}(k,t) + v^2|k|^2\Phi(k,t) = 0,
\end{equation}
which is the classical equation of motion for a harmonic
oscillator with the dispersion relation $\omega_k = v|k|$. In
analogy with the ordinary harmonic oscillator it is convenient to
work with the creation and annihilation operators $a$ and
$a^\dagger$ instead of charges and phases. However, since we are
dealing with a field equation each Fourier component is treated
independently and we write\cite{PeskinSchroeder}
\begin{eqnarray}
\Phi(k,t) &=& \sqrt{\frac{\hbar}{2C_T\omega_k}}\Big(a_k(t) +
a^\dagger_{-k}(t)\Big),
\end{eqnarray}
where $a$ and $a^\dagger$ obey the canonical commutation relations
$[a_k,a^\dagger_{k'}] = \delta(k-k')$ and $ [a_k,a_{k'}] = 0 $,
which give the statistics of the transmission line excitations
(photons).
The harmonic oscillator nature of the transmission line makes the
phase of the creation and annihilation operators rotate with
angular frequency $\omega_k$, $ a_k(t) = a_ke^{-i\omega_kt}$,
$a_k^\dagger(t) = a_k^\dagger e^{i\omega_k t}$ and we arrive at
the final expression for $\Phi$  in terms of $a_k$ and
$a^\dagger_k$
\begin{eqnarray}\label{eq:fourierexpansion}
\Phi(x,t) &=&  \sqrt{\frac{\hbar}{4\pi C_T}}\int_{-\infty}^\infty
\frac{dk}{\sqrt{\omega_k}}\left( a_ke^{-i(\omega_k t - kx)} +
a_k^\dagger e^{i(\omega_k t - kx)}\right) .
\end{eqnarray}
In our calculations, we use the in-(out) formalism introduced in
Eq. (\ref{eq:KG_sol}), to derive the circuit scattering matrix.
For this purpose it is convenient to split the integral in Eq.
(\ref{eq:fourierexpansion}) into positive and negative
wavevectors, which at $x=0$ can be recognized as the in and
out-fields of Eq. (\ref{eq:KG_sol}). As a final remark, one may
also write the fields in terms of integrals over frequency giving
the normalization
\begin{eqnarray}
\label{freq_field_def_eq}
\Phi(x=0,t) &=& \sqrt{\frac{\hbar
Z_0}{4\pi}} \int_{-\infty}^\infty
\frac{d\omega}{\sqrt{\omega}}\left[a_\omega e^{-i\omega t} +
(a_\omega)^\dagger e^{i\omega t} \right],
\end{eqnarray}
where the annihilation and creation operators in frequency space
obey the canonical commutation relations
$[a_\omega,a^\dagger_{\omega'}] = \delta(\omega-\omega')$ and
$[a_\omega,a_{\omega'}] = 0$.

\subsection{Equations of Motion: Charge Qubit}
Considering the adiabatic Hamiltoninan in Eq. (\ref{eq:H_charge})
and the relation in Eq. (\ref{eq:diff:kg_sol}), the Heisenberg
equations of motion for the circuit are
\begin{eqnarray}\label{eq:eqnsofMotionCircuit}
\dot{p_C} &=& -\frac{1}{L}\Phi_C ,\qquad \dot{p}_p =
-\sqrt{\frac{C_T}{L_T}}\left(  \frac{\partial \Phi_p}{\partial t}
- 2 \frac{\partial \Phi^{in}_p}{\partial t}\right), \nonumber \\
\dot{\Phi}_C &=& (D_{23}+g\sigma_z)(p_C+ p_p), \qquad \dot{\Phi}_p
= (D_{33}+g\sigma_z) p_p +
(D_{23}+g\sigma_z) p . \nonumber \\
\end{eqnarray}
This set of four coupled equations can easily be reduced to two,
only containing the phase operators of the oscillator and the
transmission line. Since the effective circuit only contains
linear elements it is straightforward to solve Eq.
(\ref{eq:eqnsofMotionCircuit}) in Fourier space. The linearity is
a direct consequence of the weak coupling between the oscillator
and the qubit, where we have replaced the non-linear Josephson
element with a linear capacitance. Both the outgoing field and the
charge operator coupling to the qubit are expressed in terms of
the incoming field
\begin{eqnarray}\label{eq:scattMatCharge}
N^{g/e}(\omega) &=& (D^{g/e}_{23})^2+(D^{g/e}_{33}-i Z_0 \omega)(L\omega^2-D^{g/e}_{23}), \nonumber \\
p^{g/e}_C(\omega)+p^{g/e}_p(\omega) &=&
\frac{-2iL\omega^3}{N^{g/e}(\omega)}  \Phi_{in}(\omega) = \chi_{g/e}(\omega) \Phi_{in}(\omega), \nonumber \\
\Phi_{out}^{g/e}(\omega) & = &
\frac{N^{g/e}(-\omega)}{N^{g/e}(\omega)}\Phi_{in}(\omega)
=e^{i\varphi^{g/e}(\omega)}\Phi_{in}(\omega),
\end{eqnarray}
where $Z_0 = \sqrt{L_T/C_T}$ is the characteristic impedance of
the transmission line. We also introduce the short-hand notation
$D_i^{g/e} \equiv D_i + g\sigma_z$. Since there is no dissipation
in the lumped circuit we have
$|\Phi^{out}(\omega)|=|\Phi^{in}(\omega)|$ and the in- and
outgoing field only differ by a state-dependent phase. In the
language of Subs. \ref{sec:createannihilate} we write Eqs.
(\ref{eq:scattMatCharge}) in the time domain
\begin{eqnarray}
\phi^{g/e}_{out}(x,t) &=& \sqrt{\frac{\hbar Z_0}{4\pi}}\int_0^\infty
\frac{d\omega}{\sqrt{\omega}}  \left( e^{i\phi^{g/e}(\omega)}
a_{in}(\omega)e^{-i\omega(t-x/v)}+ {\rm h.c.} \right), \nonumber \\
p^{g/e}(t )   &=& \sqrt{\frac{\hbar Z_0}{4\pi}}\int_0^\infty
\frac{d\omega}{\sqrt{\omega}}  \left( \chi_{g/e}(\omega)
a_{in}(\omega)e^{-i\omega(t-x/v)}+ {\rm h.c.} \right),
\end{eqnarray}
where the operators contributing to the charge on the oscillator
are written as one operator $p^{g/e}(t ) \equiv
p^{g/e}_C(t)+p^{g/e}_p(t)$. We now have the exact dynamics for the
combined circuit + transmission line system. However, to get
simple analytical expressions for the measurement and dephasing
time we need to write the relations $N(\omega)$ and $\chi(\omega)$
on a more simple form. We thus make a Breit-Wigner approximation,
see \ref{app:BW} Thus we write $\chi(\omega)$ as
\begin{eqnarray}\label{eq:BW_charge}
\chi^{g/e}(\omega)&=&
\frac{\chi_0^{g/e}}{1+i2Q^{g/e}\left(\frac{\omega-\omega^{g/e}_0}
{\omega^{g/e}_0}\right)} \qquad \chi_0^{g/e}=\frac{-2}{C_c Z_0
D_{23}^{g/e}}, \nonumber \\
\omega_0^{g/e} & = & \sqrt{\frac{D_{23}^{g/e}}{L(1+C_c
D_{23}^{g/e})}} \qquad Q^{g/e}= \frac{1}{Z_0 C_c^2 L
(\omega_0^{g/e})^3},
\end{eqnarray}
where the peak value $\chi_0^{g/e}$ is evaluated at the resonance
frequency $\omega = \omega_0^{g/e}$, and $Q^{g/e}$ is the quality
factor of the oscillator.

\subsection{Equations of Motion: Flux Qubit}
The flux-qubit is treated in exact analogy with the charge qubit.
The only major difference is that the qubit degree of freedom now
couples to the phase, and not charge operator of the oscillator.
Given the Hamiltonian in Eq. (\ref{eq:H_flux}) we solve for $\Phi$
and $\Phi_p^{out}$ in terms of the in-field $\Phi_p^{in}$
\begin{eqnarray}\label{eq:scattMatFlux}
N^{g/e}(\omega)&=& C(1-iC_c Z_0 \omega)(1-L^{g/e}C\omega^2)-C_c
\nonumber \\
\Phi(\omega)&=& \frac{-2 C_c C L^{g/e} \omega^2}{N^{g/e}(\omega)}
\Phi_p^{in}(\omega) \equiv \chi(\omega) \Phi_p^{in}(\omega) \nonumber \\
\Phi_p^{out}(\omega)&=&\frac{N^{g/e}(-\omega)}{N^{g/e}(\omega)}
\Phi_p^{in}(\omega) = e^{i\varphi^{g/e}(\omega)}\Phi_{in}(\omega),
\end{eqnarray}
where $\chi(\omega)$ for the flux qubit can be written on the
exact same Breit-Wigner form as in Eq.~(\ref{eq:BW_charge}). The
only difference lies in the parameters describing the peak height,
width and resonance frequency. In this case they are
\begin{equation}\label{eq:BW_flux}
\chi_0^{g/e} = -i\frac{2C_c Q^{g/e}}{C},
\qquad
\omega_0^{g/e}=\frac{1}{\sqrt{L^{g/e}C_{\rm eff}}},
\qquad
Q^{g/e}=\frac{C^2}{Z_0
C_c^2}\sqrt{\frac{L^{g/e}}{C_{\rm eff}}},
\end{equation}
where $C_{\rm eff} = C/(1 - C_c/C)$.

\subsection{Phase of the Reflected Signal}
We now have a full quantum description of oscillator and qubit
dynamics in terms of the incoming field. The interaction can be
written on a simple form and we can use this to derive simple
analytical expressions for the quantum efficiency for the
different qubits. Using the Breit-Wigner approximation, the
reflected phase has the following simple form
\begin{equation}
\varphi^{g/e}_r=
-2\arctan{\left(\frac{2Q^{g/e}(\omega-\omega^{g/e}_0)}
{\omega^{g/e}_0}\right)} .
\end{equation}
The Breit-Wigner approximation rely on the presence of a relative
pronounced resonance in the function $\chi(\omega)$, implying an
oscillator with a finite quality factor. We emphasize however that
this need not to be the case. Numerical calculations, using the
exact form for $N(\omega)$ and $\chi(\omega)$ also give a quantum
efficiency $\eta \simeq 1$. In Sec. \ref{sec:noOsc} we calculate
$\eta$ without the presence of the oscillator, thereby taking away
the resonance altogether. Even in this case, we find a quantum
efficiency of unity.

\section{DETECTING THE QUBIT STATE BY HOMODYNE DETECTION}\label{sec:readout}
The interaction between qubit and oscillator allows for an
indirect measurement of the qubit state. Since the interaction is
purely reactive and no dissipation is present in the process, all
information about the qubit is contained in the phase of the
reflected signal. To analyze the quantum efficiency of such a
read-out we use standard techniques from quantum optics and
describe it as a homodyne measurement\cite{zoller}. In this setup,
the reflected signal from the circuit is mixed with a phase-locked
local oscillator field using a linear mixer, a beam-splitter
combined with a photon detector in the optical case. By detecting
the power of this mixed signal, information about the phase can be
extracted. Labelling the mixed signal with the operator $b$ gives
the field intensity at the detector
\begin{equation}\label{eq:intensity}
N(T) = \int_0^T dt b^\dagger(t)b(t).
\end{equation}
Detecting intensity is equivalent to detecting photons, a
destructive process where photons are converted into current by a
photoamplifier. Our signal will thus be given by the annihilation
operator at the detector
\begin{equation}
b(t) = r\alpha(t) + rv(t) + t a(t),
\end{equation}
which contains the average $\alpha$ and residual $v$ of the LO
mode. The reflected signal from the circuit is given by the
operator $a$. The reflection and transmission coefficients $r$ and
$t$ are assumed to be real. For optimal efficiency we need that
$r\ll1$ which allows almost all the signal from the circuit to be
transmitted to the detector. By amplifying the local oscillator
field $r \alpha(t)$ can still be made dominant so that second
order contributions in $v(t)$ and $a(t)$ can be neglected in Eq.
(\ref{eq:intensity}). Both the LO-field and signal are modelled as
a Glauber state\cite{barnett}
\begin{equation}\label{eq:glauber}
|\{\alpha(\omega)\}\rangle = \exp{\left(\int d\omega
[\alpha(\omega)(a^{in}_\omega)^\dag-\alpha^*(\omega)a^{in}_\omega]\right)|0\rangle},
\end{equation}
where $\alpha(\omega)$ is the Fourier-transform of our drive
signal, and $|0\rangle$ is the continuum vacuum field
$a_\omega|0\rangle=0$. The drive source is nearly monochromatic
and has a narrow bandwidth distribution ($\Gamma_d \ll \omega_d$)
which is Gaussian in frequency
\begin{equation}\label{eq:freqdist}
\alpha(\omega)=\alpha_0 \frac{\omega_d}{\Gamma_d}
\frac{e^{-(\omega-\omega_d)^2/2\Gamma_d^2}}{\sqrt{\omega}} ,
\end{equation}
where $\alpha_0$ is a dimensionless constant and $\omega_d$ is the
drive frequency. At the detector we
observe the quantum mechanical expectation
value of the operator $N$.
Assuming that the deviation $v$ from
the average LO signal is in a vacuum state, this is given by
\begin{equation}
\langle N(T) \rangle  = T\big(r^2\alpha_{LO}^2 + 2tr\alpha_{LO}
\cos(\phi + \theta) \big),
\end{equation}
where $\phi$ and $\theta$ are the phases of the signal and
LO-field respectively. The first term only contains information
about the LO-field and does not contribute to the phase
information. We thus take the second term as the signal at the
detector $S$. The standard deviation of the photon number operator
is
\begin{equation}
\Delta N = r\alpha_{LO}\sqrt{T},
\end{equation}
which we take to be the signal noise $No$. The measurement
time $t_{ms}$ is defined as the time for which the signal to noise
ratio is one.
Choosing $\theta + (\phi_r^g+\phi_r^e)/2=\pi/2$, and $t\simeq 1$ it is
given by
\begin{equation}
t_{ms}=\frac{1}{\Gamma_{in}}\frac{1}{4
\sin^2{\left[\frac{\varphi_r^g-\varphi_r^e}{2}\right]}} ,
\end{equation}
where $\Gamma_{in}$ is rate of photons sent down to the circuit by
the drive. We now have a formula for the measurement time,
determined by the driving strength and the difference in the
reflected phase between the two qubit states. This we may compare
with the measurement induced dephasing.

\section{QUBIT DEPHASING}
The dephasing rate $\Gamma_\varphi$ is defined as the rate with
which the qubit looses its phase coherence. That is how fast the
off-diagonal matrix elements in the reduced qubit density matrix
go to zero. Since the phase coherence is the heart of all quantum
computation, the qubit is useless after a time given by the
dephasing time $1/\Gamma_\varphi$. The short dephasing time is the
major bottleneck preventing the progress in solid state quantum
computing. The reason for such short life-times are the ever
present strong couplings between qubit and environment inducing
all kinds of fluctuations in the qubit. We must therefore see to
that all additional sources of decoherence are minimized to reduce
this effect. One such source is the transmission line that we have
attached to the qubit for read-out. Photons are the elementary
excitations of energy in the transmission line and fluctuations in
these will dephase the qubit. In this section we calculate the
dephasing rate for two different cases. In the first case the
state of the in-field is set to a Glauber state as was done in
Sec. \ref{sec:readout} to simulate the field generated by a
monochromatic source used for readout. In the second case a
thermal distribution of photons is used to simulate the dephasing
due to the small but finite temperature in the experimental setup,
which occurs also when not reading out.

\subsection{Dephasing from Field Correlators}
We begin by considering a general pure state characterizing the
qubit and oscillator
\begin{equation}\label{eq:QbState}
|\Psi\rangle = a |g\rangle \otimes
 |\psi\rangle^{\rm osc}_g + b |e\rangle\otimes
 |\psi\rangle^{\rm osc}_e,
\end{equation}
where $a$ and $b$ are complex numbers fulfilling $|a|^2+|b|^2=1$.
At the degeneracy point the Hamiltonian describing the qubit and
interaction with the oscillator is given by
\begin{equation}
H = \frac{E_J}{2}\sigma_z + \frac{g}{2}\sigma_z P^2 =
\underbrace{\frac{E_J}{2}\sigma_z +
\frac{g}{2}\sigma_z \langle P^2\rangle}_{H_0} + \frac{g}{2}\sigma_z
p^2,
\end{equation}
where we incorporate the average value of $p$ into the Hamiltonian
$H_0$. Our first task is to calculate the time evolution of the
$\sigma_+$ operator. We use the charge qubit as example, but both
the formalism and final result apply to the flux qubit as well.
For our purposes it is convenient to make a unitary transformation
to the rotating frame of the unperturbed qubit system $H_0$
\begin{equation}
O_I(t) = e^{-\frac{i}{\hbar}\int H_0
dt}O_H(t)e^{\frac{i}{\hbar}\int H_0 dt },
\end{equation}
where $H_0$ is the unperturbed Hamiltonian. In this
frame the equation of motion for the $\sigma_+$-operator is
\begin{eqnarray}\label{eq:sigma+}
\frac{\partial \sigma_+}{\partial t} &=& -i
\frac{g}{\hbar}p^2(t)\sigma_+(0) - \frac{g^2}{\hbar^2}\int_{0}^t
\left[\sigma_+(t)p^2(t')p^2(t) + p^2(t)\sigma_+(t)p^2(t') + \right. \nonumber \\
& & \left. +p^2(t')\sigma_+(t)p^2(t) +p^2(t)p^2(t')\sigma_+(t)
\right] dt',
\end{eqnarray}
to second order in $g$. Moreover, the evolution of $\sigma_+$ is
assumed to be slow on the time-scale of the oscillator. The
dephasing is now given by multiplying Eq. (\ref{eq:sigma+}) by the
state in Eq. (\ref{eq:QbState}) from the left and right and
tracing out the oscillator degrees of freedom, giving an equation
of motion for $ab^\ast$, the off-diagonal element in the reduced
density matrix
\begin{eqnarray}\label{eq:dephasingeqn}
\frac{\partial(ab^\ast)}{\partial t} &=& -ab^\ast
\frac{g^2}{4\hbar^2}\int_{0}^t \left[ \langle
p_g^2(t')p_g^2(t)\rangle+\langle p_e^2(t)p_g^2(t')\rangle +
\right. \nonumber \\
& & \left. + \langle p_e^2(t')p_g^2(t\rangle)+\langle
p_e^2(t)p_e^2(t')\rangle\right] dt' .
\end{eqnarray}
The index on the $p^2$-operator indicates the state of the field.
We now have an expression for the dephasing rate. The only
remaining task is to put the in-field in a given state and
calculate the two-time correlators of Eq. (\ref{eq:dephasingeqn}).

\subsection{Dephasing Induced by the Read-Out}\label{sec:dephase_RO}

In the read-out process the oscillator state in Eq.
(\ref{eq:QbState}) is a coherent Glauber state, see Eq.
(\ref{eq:glauber})
\begin{eqnarray}|\Psi\rangle = a |g\rangle \otimes
 |\alpha(\omega)\rangle_g + b |e\rangle\otimes
 |\alpha(\omega)\rangle_e .
\end{eqnarray}
The oscillator state is now an eigenstate of the annihilation
operator which makes the calculation of the correlators in Eq.
(\ref{eq:dephasingeqn}) a matter of putting the creation and
annihilation operators in normal order. Using the narrow bandwidth
of the drive as given in Eq. (\ref{eq:freqdist}) the correlators
are found to be
\begin{eqnarray}
\langle p_\alpha^2(t')p_\beta^2(t)\rangle = 4\langle p_\alpha(t)
\rangle\langle p_\beta(t') \rangle\frac{\hbar
Z_0}{4\pi}\int_0^\infty
\frac{\chi_\alpha(\omega)\chi^\ast_\beta(\omega)}{\omega}
e^{-i\omega(t-t')}
+\nonumber \\
+ 2\frac{\hbar^2Z_0^2}{16\pi^2}\int_0^\infty \int_0^\infty
d\omega_1
d\omega_2\frac{\chi_\alpha(\omega_1)\chi^\ast_\beta(\omega_1)}
{\omega_1}\frac{\chi_\alpha(\omega_2)\chi^\ast_\beta(\omega_2)}{\omega_2}
e^{-i(\omega_1 + \omega_2)(t-t')}.
\end{eqnarray}
The dephasing rate is given by taking the zero frequency Fourier
transform. We make the variable transformation $t-t' = \tau$ and
extend the upper limit of the integral to infinity, an
approximation which is valid if we limit ourselves to times longer
than the field correlation time. Neglecting the imaginary part of
the transform, which only gives a small renormalization to the
qubit energy splitting, we get the simple expression
\begin{eqnarray}
Re\left\{\int_{0}^\infty \langle p_\alpha^2(t')p_\beta^2(t)\rangle
dt' \right\} = \Gamma_{in}\frac{\hbar^2 Z_0^2}{2\omega_d}
\frac{|\chi_\alpha(\omega_d)|^2|\chi_\beta(\omega_d)|^2}{\omega_d}.
\end{eqnarray}
where $\Gamma_{in}$ is the rate of photons sent down the
transmission line and $\omega_d$ is the drive frequency. Solving
Eq. (\ref{eq:dephasingeqn}) together with the result from above.
the off-diagonal element will exhibit an exponential decay with
the dephasing rate given by
\begin{eqnarray}
\Gamma_\varphi &=& \Gamma_{in}
\frac{g^2Z_0^2}{8\omega_d^2}\Big( |\chi_g(\omega_d)|^2 +
|\chi_e(\omega_d)|^2\Big)^2.
\end{eqnarray}
This result is valid for both charge and flux qubits, choosing the
correct expressions for coupling $g=g_C=C_Q/C_{osc}^2$ and
$g=g_L=L_Q/L^2$, as well as for the transfer function
$\chi_{g/e}(\omega)$ from Eqs.~(\ref{eq:scattMatCharge}) and
(\ref{eq:scattMatFlux}). As expected the dephasing rate increases
with the drive power. We also note that the rate is inversely
proportional to the drive frequency squared, not surprising since
the dephasing is given by the zero-frequency fluctuations in the
field.

\subsection{Finite Temperature Dephasing}
Because of the non-zero temperature in the environment, photons in
the transmission line will be thermally excited and contribute to
the dephasing of the qubit. It is desireble to reduce this effect
by an efficient design. We are therefore interested to see how
this dephasing scales with the different circuit parameters. The
state of the oscillator in Eq. (\ref{eq:QbState}) is given by a
thermal distribution of photons. The correlators in Eq.
(\ref{eq:dephasingeqn}) are characterized by
\begin{equation}
\langle a^\dag_\omega a_{\omega'} \rangle = \delta(\omega-\omega')
n(\omega),\qquad n(\omega)=\frac{1}{e^{\beta\hbar\omega}-1} ,
\end{equation}
where $n(\omega)$ is the Bose-occupation number. Given this it is
straightforward to calculate the dephasing rate as was done in
Subs. \ref{sec:dephase_RO}. Following Eq. (\ref{eq:dephasingeqn})
we calculate the correlator
\begin{eqnarray}\label{eq:p2corr_temp}
& & \langle p_\alpha^2(t)p_\beta^2(t') \rangle  =  \langle
p_\alpha^2 \rangle \langle p_\beta^2 \rangle + 2 \frac{\hbar^2
Z_0^2}{16\pi^2} \int_0^\infty \int_0^\infty d\omega
d\omega' \times\nonumber \\
&\times&\left[\frac{\chi_\alpha(\omega)\chi_\beta^\ast(\omega)}{\omega}
  (n(\omega)+1) e^{-i\omega(t-t')}+ \frac{\chi_\alpha^\ast(\omega)\chi_\beta(\omega)}{\omega}n(\omega)
e^{i\omega(t-t')}\right] \times \nonumber \\
&\times&\left[\frac{\chi_\alpha(\omega')\chi_\beta^\ast(\omega')}{\omega'}
(n(\omega')+1)
e^{-i\omega'(t-t')}+\frac{\chi_\alpha^\ast(\omega')\chi_\beta(\omega')}{\omega'}n(\omega')
e^{i\omega'(t-t')}\right], \nonumber \\
\end{eqnarray}
where the first term vanishes since the $p^2$-operator has zero
average. The dephasing rate is given by the Fourier transform of
Eq.~(\ref{eq:p2corr_temp}). As in Subs. \ref{sec:dephase_RO} we
make the substitution $\tau = t - t'$ and ignore the imaginary
part of the integral, giving
\begin{equation}
\Gamma_\varphi = \frac{Z_0^2g^2}{16\pi}\int_0^\infty \Big(
|\chi_g(\omega)|^2 +|\chi_e(\omega)|^2   \Big)^2 \big( n(\omega) + 1
\big)n(\omega) \frac{d\omega}{\omega^2},
\end{equation}
valid for both charge and flux qubits. Analytically we may proceed
in the Breit-Wigner approximation of the transfer functions
$\chi_{g/e}(\omega)$. We want to compare this dephasing rate with
other dephasing mechanism affecting the qubit. For this purpose it
is sufficient to make the approximation $|\chi_g|\approx|\chi_e|$,
and ignore effects of the quantum capacitance. Given that the rest
of the integrand varies slowly compared to $\chi(\omega)$ we
arrive at the following expression for the dephasing rate for the
charge qubit
\begin{equation}
\Gamma_\varphi = \omega_0Q\frac{C_Q^2}{C_{osc}^2}\big( n(\omega_0) + 1
\big)n(\omega_0),
\end{equation}
where $\omega_0 = 1/\sqrt{LC_{osc}}$ and $Q = 1/(Z_0C_c^2L
\omega_0)$ is the approximate resonance frequency and Q value. A
similar expression was found by Bertet {\it et. al.} considering a
flux qubit coupled to a SQUID, in the high Q
regime\cite{DelftFluxPhotonDephasing}. As expected the dephasing
rate scales badly with the temperature due to the bosonic nature
of the transmission line excitations. However we see that a low
oscillator $Q$-value protects the qubit from dephasing, basically
by pushing the noise spectrum up to higher frequencies.

\section{QUANTUM EFFICIENCY}\label{sec:QuantEff}
There is always a trade-off between speed in read-out and how fast
the qubit is dephased. It is impossible to obtain information
about the state of the qubit faster than the phase coherence
between the states are lost. Given a longitudinal coupling between
qubit and oscillator and a probing field in a coherent Glauber
state, which is narrow in frequency distribution, we have derived
the following general expressions for the dephasing and
measurement rate
\begin{eqnarray}\label{eq:rates}
\Gamma_\varphi &=& \Gamma_{in}\frac{(gZ_0)^2}{8\omega_d^2}\Big(
|\chi_g(\omega)|^2 + |\chi_e(\omega)|^2 \Big)^2, \nonumber \\
t_{ms}^{-1} &=& 4\Gamma_{in}\sin^2\left(
\frac{\delta\phi}{2}\right),
\end{eqnarray}
where $g$ is the coupling between qubit and oscillator degrees of
freedom and $\chi(\omega)$ describes the relation between charge
(phase) of the oscillator and the in field. We now assume that the
function $\chi(\omega)$ can be written on Breit-Wigner form as in
Eq. (\ref{eq:BW_charge}). Using this, we get explicit expressions
for the dephasing rate in terms of circuit parameters
\begin{eqnarray}
\Gamma &=& \Gamma_{in}\frac{(gZ_0)^2}{8\omega_d^2}\left(
\frac{(\chi_0^g)^2}{1 + 4 \left[\frac{Q^g}{\omega_0^g}(\omega_0^{g} - \omega)\right]^2} +
\frac{(\chi_0^e)^2}{1 + 4 \left[\frac{Q^e}{\omega_0^e}(\omega_0^{e} - \omega)\right]^2 }
\right)^2
\end{eqnarray}
and the phase difference between the two qubit states gets the simple
form
\begin{equation}
\delta\phi = 2\arctan\left( \frac{2Q^g}{\omega_0^g}(\omega
-\omega_0^g)\right)-
              2\arctan\left( \frac{2Q^e}{\omega_0^e}(\omega -\omega_0^e)\right).
\end{equation}
Choosing the drive frequency to be centered between the two resonance frequencies $\omega_d = \frac{\omega_0^g + \omega_0^e}{2}$,
these expressions simplify to
\begin{eqnarray}\label{eq:rateBW}
\Gamma_\varphi &=&
\Gamma_{in}\frac{(gZ_0)^2}{8\omega_d^2} \left[
\frac{(\chi_0^g)^2}{x_g(x_g^{-1} + x_g) }  +
\frac{(\chi_0^e)^2}{x_e(x_e^{-1} + x_e) }
\right]^2   \nonumber \\
t_{ms}^{-1} &=& 4\Gamma_{in} \sin^2\Big( \arctan x_e +\arctan x_g
\Big),
\end{eqnarray}
where $x_{g/e} = Q^{g/e}(\omega_0^e -\omega_0^g )/\omega_0^{g/e}$. Defining the quantum efficiency as
$\eta = \Gamma_{ms}/(2\Gamma_\phi)$
and using Eq. (\ref{eq:rateBW}) the efficiency is given in terms
of general parameters in the Breit-Wigner approximation. Using the
parameters given for the charge and phase qubit in Eqs.
(\ref{eq:BW_charge}) and (\ref{eq:BW_flux}) respectively and
neglecting second order effects in the qubit-oscillator coupling
(small parameter is $C_Q^2/C_{osc}^2$ and $L_Q^2/L^2$
respectively) we get a quantum efficiency $\eta = 1$ for both
charge and flux qubit.

For another choice of drive frequency, we find an efficiency below
unity. For these non-optimal frequencies, the charge magnitude on
the oscillator differ for the two different qubit states. In this
case, since the time to build up a charge on the oscillator depend
on the qubit state, the time response of the oscillator will
differ. Thus some information of the qubit state will be encoded
in the response time, and since our setup only detects the phase,
it misses this information, making the efficiency decrease. For
the optimal drive frequency above, the charge on the oscillator is
the same regardless of qubit state, and all information is encoded
in the signal phase.

In a more realistic situation, the qubit will of course be subject
to various additional sources of decoherence, i.e. thermal photons
in the read-out lines and charge fluctuations in the environment,
and these will reduce the quantum efficiency. Taking these sources
into account is however beyond the scope of this paper.

\section{MEASURING THE QUANTUM CAPACITANCE WITHOUT OSCILLATOR}\label{sec:noOsc}
\begin{figure}[!ht]
\psfrag{Vg}[][][0.9]{$V_g$}
\psfrag{phi_g}[][][0.9]{$$}
\psfrag{C_g}[][][0.9]{$C_g$}

\psfrag{phi_m}[][][0.9]{$$}
\psfrag{C_m}[][][0.9]{$C_m$}

\psfrag{phi_J}[][][0.9]{$$}
\psfrag{C_J}[][][0.9]{$C_J$}
\psfrag{E_J}[][][0.9]{$E_J$}

\psfrag{phi_l}[][][0.9]{$$}
\psfrag{L}[][][0.9]{$L$}

\psfrag{phi_C}[][][0.9]{$$}
\psfrag{C}[][][0.9]{$C$}

\psfrag{phi_in}[][][0.9]{$$}
\psfrag{C_c}[][][0.9]{$C_{in}$}

\psfrag{phi1}[][][0.9]{$$}
\psfrag{phi2}[][][0.9]{$$}
\psfrag{C_c}[][][0.9]{$C_c$}
\psfrag{C_T}[][][0.9]{$C_T\Delta x$}
\psfrag{L_T}[][][0.9]{$L_T\Delta x$}

\psfrag{deltaphi}[][][0.9]{$$}

\psfrag{PHI}[][][1.5]{$\Phi$}
\psfrag{M}[][][0.9]{$M$}
\psfrag{L_J}[][][0.9]{$L_J$}
\psfrag{ChargeQubit}[][][0.9]{Charge Qubit}
\psfrag{Oscillator}[][][0.9]{Oscillator}
\psfrag{TransmissionLine}[][][0.9]{Transmisson Line}

\includegraphics[width=12cm]{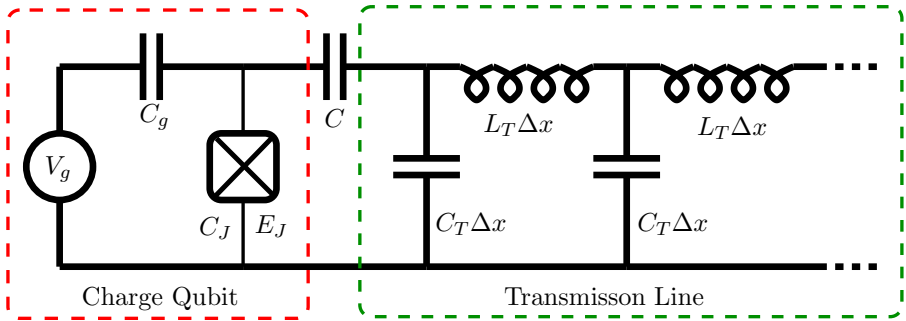}
\caption{Cooper-pair box operated as a charge qubit. The read-out
  circuit only consists of a measurement capacitance $C$, still giving
  a full quantum efficiency $\eta = 1$.}
\label{fig:noOsc}
\end{figure}

The calculation of the quantum efficiency for the charge and flux
qubit was done assuming a rather sharp resonance in the function
$\chi(\omega)$ approximating it with a Breit-Wigner form. We can
however show that the result still holds if we remove the
oscillator, thereby taking away the resonance completely. For this
purpose we consider the circuit in Fig. \ref{fig:noOsc}
where the charge qubit is directly coupled to the transmission line via
the capacitor $C$. For this circuit the Hamiltonian is given by
\begin{equation}
H = \frac{p_1^2}{2C_{\rm g/e}} + \frac{1}{\Delta x} \sum_i \left( \frac{
(p_{(i+1)}^p)^2}{2C_T} + \frac{(\Phi^p_{i+1} - \Phi^p_{i}
)^2}{2L_T}\right),
\end{equation}
where $C_{\rm g/e} = C/(1 + gC \sigma_z)$ and $g = (8E_C^2)/(e^2E_J)$
is the coupling between qubit and transmission line. The functions
relating the in-field with the charge at the transmission-line
boundary and the reflected signal are
\begin{equation}
p(\omega) = \frac{2\omega C_{\rm g/e} Z_C(\omega)} {i Z_C(\omega)
+
  Z_0} \Phi^{in}(\omega) \equiv \chi(\omega) \Phi^{in}(\omega), \
  \Phi^{out}(\omega)=\frac{i Z_C(\omega)-Z_0}{i Z_C(\omega)+Z_0}
\Phi^{in}(\omega),
\end{equation}
where $Z_C(\omega) = 1/(\omega C_{\rm g/e})$ is the impedance of
the effective qubit-capacitor circuit. As in Sec.
\ref{sec:QuantEff} we use a narrow bandwidth drive signal with
drive frequency $\omega_d$ and Eq. (\ref{eq:rates}) to calculate
the dephasing and measurement rate
\begin{eqnarray}
\Gamma_\varphi &=& \Gamma_{in}\frac{(gZ_0)^2}{8\omega_d^2}\left(
  \frac{4(\omega_d
  C_g)^2}{1 + (\omega_d C_g Z_0)^2} + \frac{4(\omega_d
  C_e)^2}{1 + (\omega_d C_e Z_0)^2}
 \right)^2, \nonumber \\
t_{ms}^{-1} &=& 4\Gamma_{in} \sin^2\Big(\arctan{(\omega_d Z_0
C_e)} -\arctan{(\omega_d Z_0 C_g)}  \Big).
\end{eqnarray}
In this case a quantum efficiency of unity is also obtained,
neglecting second order effects in the qubit-transmission line
coupling $g$. This is however of more academic interest
considering that typical parameters $Z_0=50$ Ohm, $C_g-C_e=1$ fF,
give a very small phase-difference on the order of $\delta\varphi
\sim 10^{-4}$, even for a high drive frequency of $\omega_d=1$
GHz.  A too small phase difference implies a small signal, which
in a real setup will drown in the amplifier noise.

\section{CONCLUSION}
We have described dispersive charge and flux qubit readout using a
slow oscillator as a quantum measurement process. Our main
approximation is that the qubit follows the oscillator
adiabatically, and we give conditions for this approximation to
hold. Within this approximation the qubit acts as an effective
linear circuit element in the oscillator circuit. For the charge
qubit it is a capacitance for the flux qubit an inductance. The
quantum dynamics of the effectively linear oscillator-transmission
line is readily solved, and we get expression for all quantities
in terms of the input field of the transmission line. Especially
we treat the coupling between the oscillator and transmission line
nonperturbatively.

We use this formalism to calculate the quantum efficiency of the
readout. This is found to be unity, within our approximation, for
both charge and flux qubit readout. As an extra example we also find
unit quantum efficiency for a charge qubit directly coupled to a
transmission line.

\appendix

\section{THE LEGENDRE TRANSFORM}\label{legendre}
Given the Lagrangian as presented in Eq.
(\ref{eq:Lagrangian_vec}), and neglecting the potential part,
\begin{equation}\label{eq:B1}
L = \frac{1}{2}\dot{\phi}_i C_{ik} \dot{\phi}_k - a_k
\dot{\phi}_k,
\end{equation}
we define the conjugated momenta according to
\begin{equation}\label{eq:B2}
p_j \equiv \frac{\partial L}{\partial\dot{\phi}_j } = \dot{\phi}_i
C_{ij} -  a_j.
\end{equation}
Given this, the Hamiltonian is obtained by a Legendre
transformation of the Lagrangian
\begin{eqnarray}
H &=& p_j\dot{\phi}_j - L= p_j(p_k + a_k)(C^{-1})_{kj} -
\nonumber\\
&-& \frac{1}{2}(p_i + a_i)
\underbrace{(C^{-1})_{ij}C_{jk}}_{\delta_{ik}} (C^{-1})_{km}(p_m +
a_m) +   a_i(p_k + a_k)(C^{-1})_{ki} =  \nonumber \\
&= & p_j(p_k + a_k)(C^{-1})_{kj} - \frac{1}{2}(p_i +
  a_i)(C^{-1})_{im}(p_m + a_m) + \nonumber \\
  &+&  a_i(p_k + a_k)(C^{-1})_{ki}
=  (p_i + a_i)(p_k + a_k)(C^{-1})_{ki} - \nonumber \\
&-& \frac{1}{2}(p_i +
  a_i)(C^{-1})_{im}(p_m + a_m)
= \frac{1}{2}(p_i + a_i)(C^{-1})_{km}(p_m + a_m) , \nonumber \\
\end{eqnarray}
making the derivation of the Hamiltonian a
matter of inverting the matrix $\hat{C}$, and subtracting the
potential part of the Lagrangian.

\section{BREIT-WIGNER APPROXIMATION; CHARGE QUBIT $+$
  OSCILLATOR}\label{app:BW}
We are interested in the resonant behaviour of the
transfer/response functions around $\omega=1/\sqrt{L C_{osc}}$. We
expand the relevant solution to $N^{g/e}(\omega)=0$ to first order
in $Z_0$
\begin{eqnarray}
\omega_{res} &=& \sqrt{\frac{D_{23}^{g/e}}{L(1+C_c
D_{23}^{g/e})}}+ i Z_0 \frac{C_c (D_{23}^{g/e})^{5/2}}{(1+C_c
D_{23}^{g/e})^{3/2}\sqrt{L}} \equiv
\omega_0^{g/e}(1+i\frac{1}{2Q^{g/e}}), \nonumber \\
\omega_0^{g/e} &=& \sqrt{\frac{D_{23}^{g/e}}{L(1+C_c
D_{23}^{g/e})}},\ \ Q^{g/e}= \frac{1}{Z_0 C_c^2 L
(\omega_0^{g/e})^3},
\end{eqnarray}
leading to
\[N^{g/e}(\omega)\approx \frac{2D_{23}^{g/e}}{C_c \omega_0^{g/e}}
\left(\omega-\omega_0^{g/e} - i
\frac{\omega_0^{g/e}}{2Q^{g/e}}\right) =
-i\frac{D_{23}^{g/e}}{Q^{g/e} C_c}
\left(1+i2Q^{g/e}\frac{\omega-\omega_0^{g/e}}{\omega_0^{g/e}}\right).\]
Thus we approximate the transconductance as (assuming $C_c Z_0
\omega_0 \ll 1$)
\begin{eqnarray}
\chi^{g/e}(\omega)&=&\frac{-2}{C_c Z_0 D_{23}^{g/e}}
\frac{\chi_0^{g/e}}{1+i2Q^{g/e}
\left(\frac{\omega-\omega^{g/e}_0}{\omega^{g/e}_0}\right)},
\nonumber\\
\chi_0^{g/e}&=&\frac{-2}{C_c Z_0 D_{23}^{g/e}} \approx
-2\sqrt{\frac{Q^{g/e}}{L Z_0 \omega_0^{g/e}}} ,
\end{eqnarray}
and the reflected phase as
\begin{equation}
\varphi^{g/e}_r=2 \arg{\left[\frac{\omega^{g/e}_0}{2Q^{g/e}}-i
(\omega-\omega^{g/e}_0)\right]}=-2\arctan{\left(\frac{2Q^{g/e}(\omega-\omega^{g/e}_0)}{\omega^{g/e}_0}\right)}
.
\end{equation}

\section*{ACKNOWLEDGMENTS}
We thank Chris Wilson, Per Delsing, Tim Duty, Vitaly Shumeiko,
G{\"o}ran Wendin and Margareta Wallquist for valuable discussions.
We would also like to thank Matthias Braun for useful comments on
the manuscript. This work was supported by the Swedish SSF and VR,
and by the EU under the EUROSQIP programme.

\end{document}